\documentclass[prl,reprint,superscriptaddress]{revtex4-1}

\usepackage{graphicx,color}
\usepackage{amssymb,amsmath}

\begin{document}

\title {Two-photon double ionization of neon using an intense attosecond pulse train}

\author{B. Manschwetus}
\thanks{These authors contributed equally to this work}
\author{L. Rading}
\thanks{These authors contributed equally to this work}
\author{F. Campi}
\author{S. Maclot}
\author{H.~Coudert-Alteirac}
\author{J.~Lahl}
\author{H.~Wikmark}
\author{P. Rudawski}
\author{C. M. Heyl}
\affiliation{Department of Physics, Lund University, P.O. Box 118, 22100 Lund, Sweden}

\author{B. Farkas}
\author{T. Mohamed}
\affiliation{ELI-HU Non-Profit Ltd., Dugonics ter 13, Szeged 6720, Hungary}

\author{A. L'Huillier}
\author{P. Johnsson}
\email[E-mail: ]{per.johnsson@fysik.lth.se}
\affiliation{Department of Physics, Lund University, P.O. Box 118, 22100 Lund, Sweden}

\date{\today}

\begin{abstract}
We present the first demonstration of two-photon double ionization of neon using an intense extreme ultraviolet (XUV) attosecond pulse train (APT) in a photon energy regime where both direct and sequential mechanisms are allowed. For an APT generated through high-order harmonic generation (HHG) in argon we achieve a total pulse energy close to 1~$\mu$J, a central energy of 35~eV and a total bandwidth of $\sim30$~eV. The APT is focused by broadband optics in a neon gas target to an intensity of $3\cdot10^{12}$~W$\cdot$cm$^{-2}$. By tuning the photon energy across the threshold for the sequential process the double ionization signal can be turned on and off, indicating that the two-photon double ionization predominantly occurs through a sequential process.
The demonstrated performance opens up possibilities for future XUV-XUV pump-probe experiments with attosecond temporal resolution in a photon energy range where it is possible to unravel the dynamics behind direct vs. sequential double ionization and the associated electron correlation effects.
\end{abstract}

\maketitle

\begin{figure}
	\centering
	\includegraphics[width=0.35\textwidth]{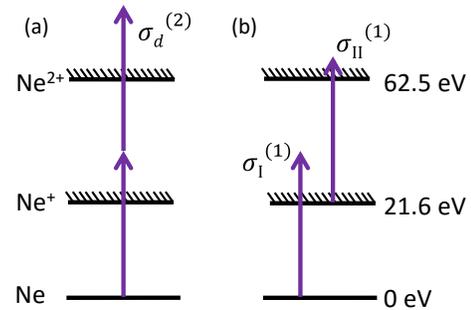}
	\caption{Possible pathways when a neon atom is doubly ionized by two photons in a direct (a) or sequential (b) process. The relevant photoionization cross sections are indicated in the figure.} 
	\label{fig:Introduction}
\end{figure}

Double photoionization of atoms or molecules can occur through the absorption of either a single energetic photon or several less energetic photons. Single-photon multiple ionization is typically studied at synchrotron facilities where the photon energies can be high but the achievable peak intensities are low. Such experiments have led to an increased understanding of electron-electron correlation and have also provided an important tool to benchmark the theory of fundamental two-electron systems like He and H$_{2}$~\cite{KnappPRL2002,AkouryScience2007}. At lower photon energies and high peak intensities, two or more photons can be used for ionization. In this regime, nonlinear processes in atoms and molecules can be studied and pump-probe experiments become possible. While multi-photon ionization using ultrashort intense infrared (IR) or visible laser pulses in the strong-field regime has been extensively studied~\cite{AgostiniIEEE1970}, very few studies have been performed in the vacuum ultraviolet (VUV) and extreme ultraviolet (XUV) regime where absorption of one or two photons is sufficient to overcome the ionization thresholds~\cite{SorokinPRA2007,PapadogiannisPRL2003,MidorikawaPiQE2008}. Double ionization by absorption of more than one photon may occur either through a direct process where the photons are absorbed simultaneously or through a sequential process where the electrons are removed one at a time from the atom or ion, as depicted in Fig.~\ref{fig:Introduction}. 
For pulse durations comparable to the time it takes for the system to relax to the ionic ground state, typically below 1 fs, the distinction between the two mechanisms becomes meaningless. The development of experimental tools with the ability to carry out time-resolved measurements at such short time scales will thus open up an intriguing regime where charge re-arrangement and electron correlation can be studied on their natural timescale.

The intensities required to induce multielectron, few-photon, processes in the VUV or XUV photon energy regime have so far mainly been available at free electron lasers (FELs), where the dependence on intensity and wavelength for multiphoton ionization and the competition between direct and sequential two-photon double ionization have been studied~\cite{SorokinPRA2007, MoshammerPRL2007, KurkaJPB2009}, with pulse durations in the 10 to 100~fs regime. During recent years, laser-driven high-order harmonic generation (HHG) sources with potential to produce pulses in the attosecond regime have started to reach the intensity levels needed for studying nonlinear phenomena. However, to date only a few HHG-based experiments have been reported where two-photon double ionization in the XUV regime was studied~\cite{TzallasNature2003, NabekawaPRL2005, TakahashiNC2013, TzallasNP2011}, since the approach still presents a formidable experimental challenge due to the inherently low conversion efficiency of the generation process. These studies use xenon as the generation medium in order to reach sufficient XUV intensities, thus limiting the highest available photon energy to regions where the direct two-photon double ionization channel dominates, since three photons are needed to access the sequential channel.

In this Letter we present the first demonstration of two-photon double ionization of neon using an attosecond pulse train (APT) generated in argon with individual pulse durations of $\approx300$~as. The APT has a total pulse energy around 1 $\mu$J and is focused to an intensity of $3\cdot10^{12}$~W$\cdot$cm$^{-2}$. With a central energy of 35~eV and a total bandwidth of $\sim30$~eV the APT covers the spectral ranges of both the direct and the sequential two-photon double ionization channels in neon. 
When tuning the photon energy below the threshold for the sequential channel the double ionization signal disappears, indicating that the two-photon double ionization predominantly occurs through a sequential process. In addition, we experimentally determine the single-photon ionization cross section for Ne$^+$ and find a good agreement with earlier measurements~\cite{CovingtonPRA2002, SorokinPRA2007}. 

Figure~\ref{fig:Introduction} shows the possible outcomes when a neon atom is doubly ionized by two photons either in a direct or in a sequential process as shown in Fig.~\ref{fig:Introduction}(a) and (b), respectively. While the direct and the sequential processes cannot be distinguished by measuring doubly charged ion yields, as they both involve absorption of two photons and depend quadratically on the intensity, they behave differently in two ways. First, for the direct channel the electrons share the total excess energy continuously and are preferentially emitted ``back-to-back."~\cite{FeistPRL2009}. For the sequential channel the electrons have discrete energies corresponding to the excess energy in each ionization step since they are removed one at a time. Despite this, a certain degree of angular correlation, possibly due to the coherent superposition of the $^2P_{3/2}$ and $^2P_{1/2}$ states of the Ne$^+$ ion, has been observed both experimentally and theoretically~\cite{FritzscheJPB2008,KurkaJPB2009}. Second, the direct and sequential ionization channels have different temporal behavior as the intermediate ionic state in the sequential channel has a long lifetime. 
Within lowest order perturbation theory, for a given photon flux, $F$~[photons$\cdot$cm$^{-2} \cdot$s$^{-1}$], and pulse duration, $\tau$, the ratio between the sequential and direct double ionization yields may be approximated as
\begin{equation}
\frac{\mathcal{N}^{2+}_s}{\mathcal{N}^{2+}_d} = \frac{ \left(\sigma^{(1)}_{\mathrm{I}} F\tau\right) \left(\sigma^{(1)}_{\mathrm{II}} F\tau\right)} {2\sigma^{(2)}_d F^2 \tau} = \frac{\sigma^{(1)}_{\mathrm{I}} \sigma^{(1)}_{\mathrm{II}}}{2 \sigma^{(2)}_{d}} \cdot \tau
\label{eq:seqvsdir}
\end{equation}
where $\sigma^{(2)}_{d}$ is the two-photon double ionization cross section for the direct process and $\sigma^{(1)}_\mathrm{I,II}$ the one-photon single ionization cross-sections for Ne and Ne$^+$, respectively. The factor of two in the denominator is due to the time-ordering of the ionization events in the sequential process. 

As shown in equation~\ref{eq:seqvsdir}, the branching ratio between the sequential and the direct process will depend on the pulse duration. Using measured~\cite{WestPRSA1976, CovingtonPRA2002} and calculated~\cite{ForrePRL2010} values for the cross sections close to the threshold for two-photon sequential double ionization, Eq.~\ref{eq:seqvsdir} predicts that the yield of the direct process will become comparable to that of the sequential one at pulse durations below $\sim$500~as. Detailed theoretical calculations indicate that in helium the transition between the two regimes occurs for pulse durations just below 1~fs~\cite{FeistPRL2009}. Due to the requirement of attosecond pulse durations, so far two-photon direct double ionization has only been observed experimentally for photon energies below the threshold of the sequential channel, ~\cite{SorokinPRA2007, TzallasNature2003, NabekawaPRL2005, TakahashiNC2013, TzallasNP2011}.

\begin{figure}
	\centering
	\includegraphics[width=0.45\textwidth]{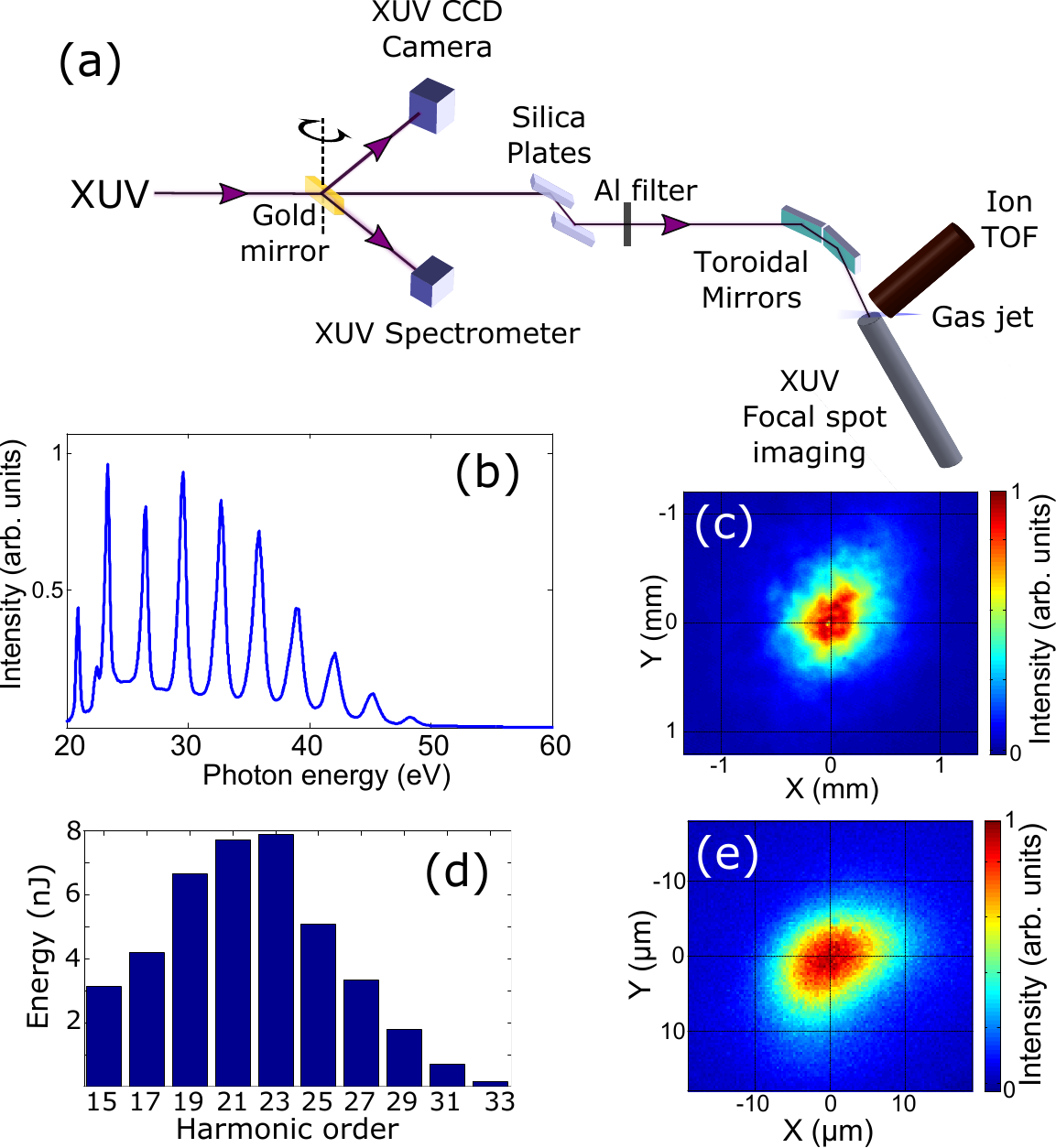}
	\caption{Experimental setup and XUV pulse characterization results for HHG in argon. (a) Schematic experimental setup. (b) Generated XUV spectrum. The peak at 20.5~eV is due to second order diffraction from the grating. (c) Far-field XUV beam profile. (d) Energy per harmonic on target. (e) XUV focal spot image. }
	\label{fig:exp_setup}
\end{figure}

The principle of the experiment as well as the characteristics of the HHG source are shown in Fig.~\ref{fig:exp_setup}, with a schematic drawing of the experimental setup in panel (a). The HHG is driven by a high-power Ti:Sapphire chirped pulse amplification laser which delivers pulses at a central wavelength of 800 nm with a temporal duration of 35~fs at 10~Hz repetition rate and with a pulse energy of up to 80~mJ after compression. The IR pulses, with a beam diameter of 30~mm, are loosely focused by a $f=9$~m lens into the generation chamber where HHG takes place in a 6~cm long static gas cell (not shown in the figure)~\cite{TakahashiPRA2002,RudawskiRSI2013}. 
After the HHG a rotatable gold mirror can be used to send the generated XUV pulses to different diagnostic devices. The spectrum, measured by a flat-field grating spectrometer, is shown in Fig.~\ref{fig:exp_setup}(b) and harmonics from order 15 (23.3~eV) up to order 33 (51.2~eV) can be seen. 
Based on the properties of the generating IR pulses, the HHG spectrum and previous APT measurements performed in our laboratory~\cite{LopezMartensPRL2005,VarjuJMO2005}, the total APT duration is estimated to 20~fs, thus containing $\approx15$~attosecond pulses with estimated individual pulse durations of $\approx300$~as.
The far-field XUV beam profile, measured with a calibrated XUV CCD camera, is shown in Fig.~\ref{fig:exp_setup}(c). The total energy above 20~eV of the generated APT in argon was estimated to $0.8\pm 0.3$~$\mu$J.

To eliminate the IR field before the experiment we use two grazing incidence silica plates, anti-reflection coated for the IR, together with a 200~nm thick Al-filter. The XUV pulses are then focused using two toroidal mirrors with a total focal length of 17~cm. The two mirrors, arranged in a Wolter configuration to minimize coma aberration~\cite{WolterADP1952}, are gold coated and designed for a grazing angle of 15$^\circ$ which allows for a theoretical reflectivity of 46\% after two reflections for the full bandwidth of the APT. By taking into account the transmission of all elements of the beamline, we estimate a total APT energy in focus of 40~nJ for generation in argon. In Fig.~\ref{fig:exp_setup}(d) the calculated energy per harmonic on target is shown. 

The focal spot is characterized by positioning a Ce:YAG scintillation crystal in the focus and imaging its surface using a long working distance optical microscope. The measured spot is shown in Fig.~\ref{fig:exp_setup}(e). It is slightly elliptical with 11~$\mu$m~x~16~$\mu$m FWHM. This is 2-3 times larger than predicted by the raytracing simulations, which we attribute to the remaining aberrations manifested in the elliptical shape of the focal spot. Using a pulse energy of 40~nJ in the focus and assuming an APT with 15 pulses with an individual duration of 300 as, we estimate the APT peak intensity in the focus to $3\cdot10^{12}$~W$\cdot$cm$^{-2}$.

\begin{figure}
	\centering
	\includegraphics[width=0.45\textwidth]{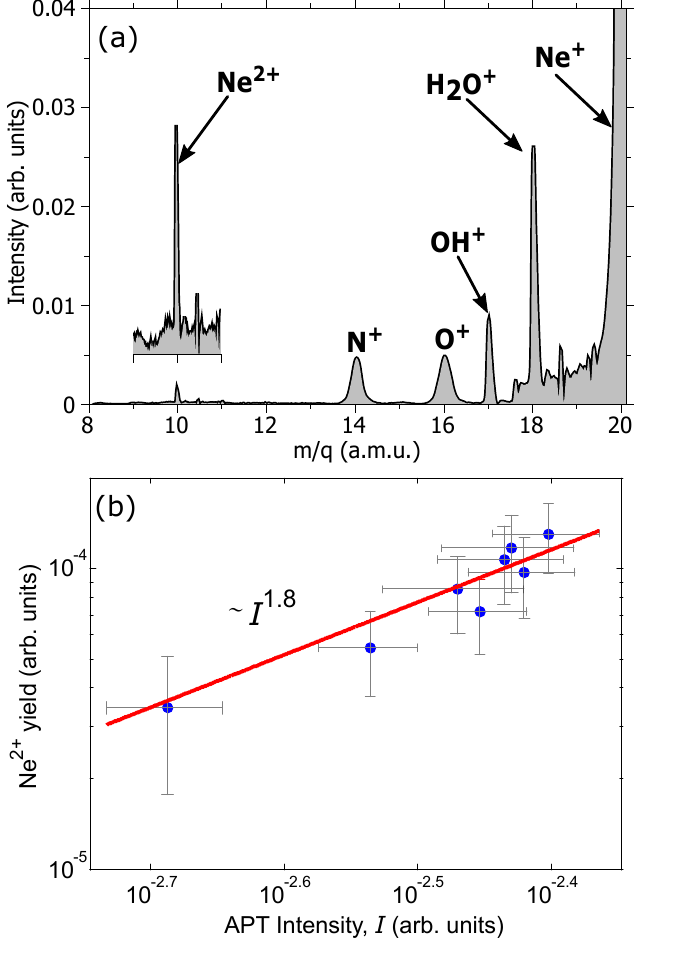}
	\caption{Two-photon double ionization of neon using an APT generated in argon. (a) Ion mass spectra. For better visibility the signal between $m/q=9$ and 11~a.m.u. is shown magnified. (b) Double logarithmic plot of the Ne$^{2+}$ yield versus the APT intensity. The error bars indicate the standard deviation of the signal after integrating over 100 shots.}
	\label{fig:nonlinearsignal}
\end{figure}

To perform the non-linear ionization experiment, the Ce:YAG crystal was replaced with an ion time-of-flight (TOF) spectrometer and a pulsed neon gas jet. The TOF spectrometer was operated in Wiley-McLaren conditions, allowing for good time-focusing over a large interaction volume \cite{WileyRSI1955}. Two micro-channel plates followed by a phosphor screen were used to detect the ions. A mass spectrum from neon is shown in Fig.~\ref{fig:nonlinearsignal}(a). Apart from the neon ionic species, the spectrum contains a few peaks due to residual gas and water contamination of the gas line. At $m/q=20$~a.m.u., the main Ne$^+$ peak is visible, although heavily saturated in the plot. The peak at $m/q=10$~a.m.u. corresponds to the doubly charged neon ion Ne$^{2+}$ with a ratio between the yields of Ne$^{2+}$ and Ne$^+$ of 0.35\%. For double ionization of neon with a single photon, photon energies larger than 62.5~eV are required, which are not available in the experimental spectrum [see Fig.~\ref{fig:exp_setup}(b)]. This implies that the observed double ionization involves the absorption of more than one photon. 

To confirm this we studied the non-linearity of the Ne$^{2+}$ yield as a function of the APT intensity. The XUV flux was adjusted by changing the gas pressure in the HHG gas cell, which had no major effect on the spectrum of the APT. To monitor the APT intensity we used the photoionization yield of H$_2$O$^+$. Since the ionization potential of H$_2$O is 12.6~eV \cite{Lias2014}, all photon energies within the experimental spectrum are able to photoionize it with a single photon and the measured yield of singly charged water ions is therefore proportional to the APT intensity. The result is shown in Fig.~\ref{fig:nonlinearsignal}(b) in a double logarithmic plot. A linear fit to the data retrieves a slope of 1.8, close to the expected slope of 2 for a two-photon process. 

\begin{figure}
	\centering
	\includegraphics[width=0.45\textwidth]{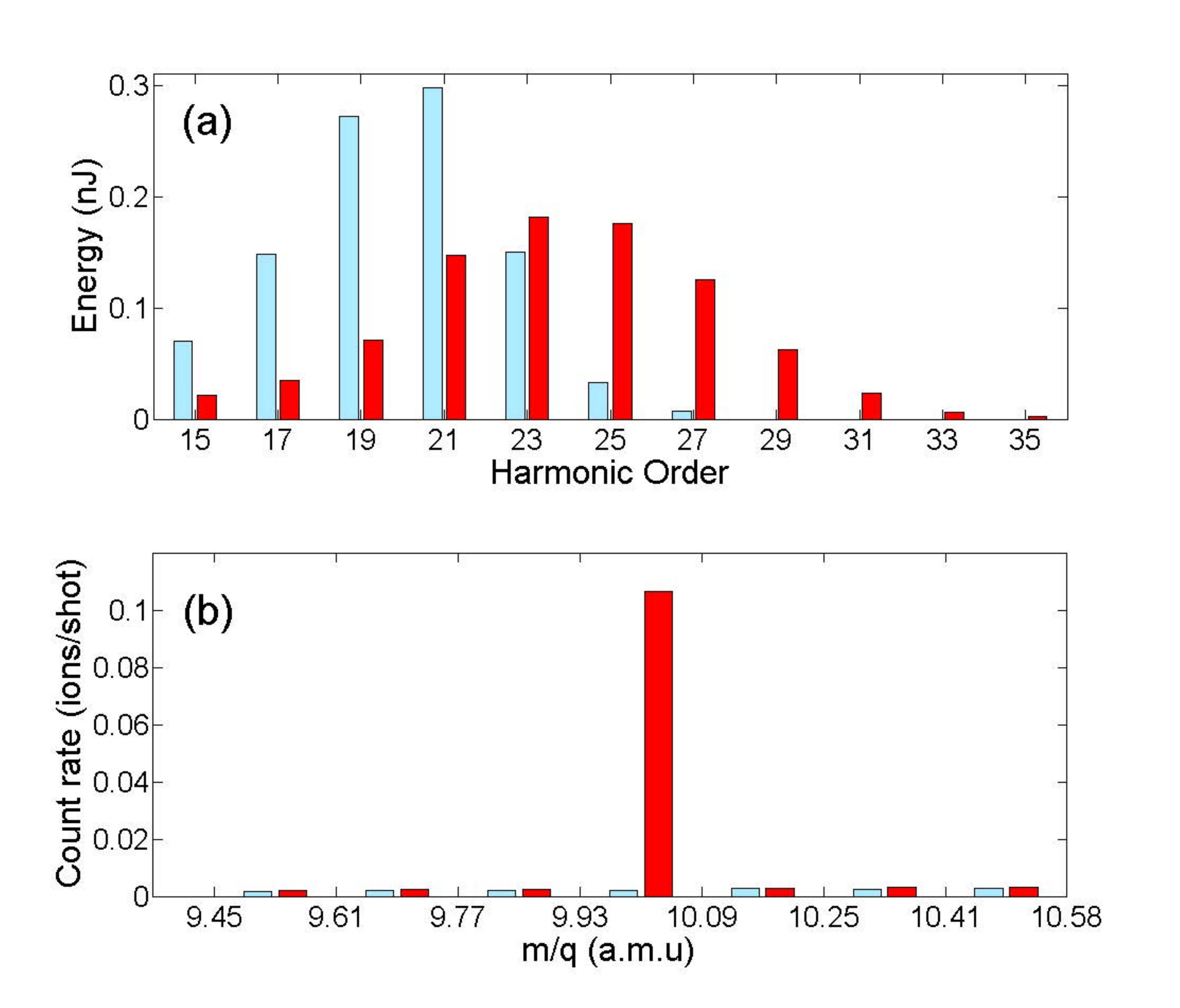}
	\caption{Ionization of neon using APTs generated in krypton. (a) The energy per harmonic on target is shown for the low (blue bars) and high (red bars) cut-off case. (b) Ion count rates in the time-of-flight region around the Ne$^{2+}$ signal ($m/q=10$) for the low (blue bars) and high (red bars) cut-off case.}
	\label{fig:krypton}
\end{figure}

The on-target APT spectrum generated in Ar includes harmonics from 15 to 33, covering the whole spectral range from 20 to 50 eV [see Fig.~\ref{fig:exp_setup}(d)], and thus both direct and sequential two-photon double ionization are possible. The APT intensity dependence of the Ne$^{2+}$ yield is quadratic for both processes and cannot be used to differentiate them.
To identify the dominant ionization pathway, we generated APTs in krypton and varied the cut-off energy across the threshold for the sequential channel (40.9~eV) while maintaining an approximately constant APT energy on target ($\approx1$~nJ). The use of Kr rather than Ar was motivated by the presence of a Cooper minimum in the Ar spectrum above the direct-sequential threshold, making difficult to identify the ``true" cutoff energy. The position of the cutoff, which is a highly nonlinear function of the IR laser intensity, was changed by varying the laser energy focused in the generation cell. The APT pulse energy was kept constant by adjusting the pressure in an absorption cell filled with argon introduced in the beamline after HHG.
The optimization of HHG for best cutoff tunability resulted in lower energy per harmonic on target than in Ar [compare Fig.~\ref{fig:exp_setup}(d) and Fig.~\ref{fig:krypton}(a)].
In order to obtain a sufficiently good signal-to-noise ratio, an ion-counting technique was applied to the single shot TOF traces around the expected time of flight for Ne$^{2+}$. Figure~\ref{fig:krypton}(b) shows the extracted ion count rates for a cut-off below (blue bars) and above (red bars) 40.9~eV. The Ne$^{2+}$ is only visible above the noise for the high-energy cut-off, indicating that the sequential channel is the dominant one in the current experiment. Using Eq.~\ref{eq:seqvsdir} and an equivalent pulse duration of $\tau=4.5$~fs by assuming 15 attosecond pulses, each with an individual duration of 300~as, we obtain a ratio of more than 20:1 between the expected ionization rates of the sequential and the direct channel, strengthening the conclusion that the sequential channel dominates, in agreement with the results of more sophisticated theoretical modeling~\cite{PalaciosPRA2009, FeistPRL2009}.

We now return to the results shown in Fig.~\ref{fig:nonlinearsignal} and the measured ratio of 0.35\% between the yields of Ne$^{2+}$ and Ne$^+$. Using rate equations and assuming a sequential process, this ratio can be approximated as
\begin{equation}
\frac{\mathcal{N}^{2+}_s}{\mathcal{N}^{+}} = \frac{\sigma^{(1)}_{\mathrm{II}}F_{\mathrm{II}}}{2 }\cdot \tau
\label{eq:doublevssingle}
\end{equation}
where $\mathcal{N}^+$ is the yield of Ne$^+$ ions and $F_{\mathrm{II}}$ is the photon flux [photons$\cdot$cm$^{-2}\cdot$s$^{-1}$] of photons with sufficiently high energy to ionize Ne$^{+}$ (harmonics of order 27 and higher). Using the measured energies per harmonic pulse from Fig.~\ref{fig:exp_setup}(d) we obtain a value for the single-photon single ionization cross section of Ne$^+$, ${\sigma^{(1)}_\mathrm{II}\approx7\cdot10^{-18}}$~cm$^2$, in good agreement with earlier experimental results measured at synchrotrons (${6\pm1\cdot10^{-18}}$~cm$^2$)~\cite{CovingtonPRA2002} and FELs (${7\pm1\cdot10^{-18}}$~cm$^2$)~\cite{SorokinPRA2007}, further supporting the conclusion that the sequential channel dominates.

In conclusion we have demonstrated, for the first time, two-photon double ionization of neon using an APT generated in argon with individual pulse durations of $\sim$300~as, a central energy of 35~eV and a total bandwidth of $\sim$30~eV covering the spectral ranges of both the direct and the sequential double two-photon ionization channels. 
By generating APTs in krypton, we were able to tune the photon energy over the threshold of the sequential channel and could conclude that, in agreement with the estimation using a simple theoretical model, despite the attosecond pulse structure of the APT, the two-photon double ionization predominantely occurs through a sequential process.  
Finally, from the ratio of the measured ion yields we determine the single photon ionization cross section for Ne$^+$ in good agreement with earlier measurements. 

Our experimental setup allows us to make use of the full bandwidth of the XUV radiation through the use of grazing incident optics for focusing and filtering of the IR, imposing no lower limit on the achievable pulse duration in terms of the available bandwidth. The demonstrated performance opens up possibilities for future XUV-XUV pump-probe experiments with attosecond temporal resolution, with a potential to unravel e.g. the dynamics behind direct vs. sequential double ionization and associated electron correlation effects~\cite{FeistPRL2009, PalaciosPRA2009}. Further, pump-probe experiments with attosecond time resolution are expected to be a useful tool for studies of charge migration in molecules, where recent theoretical~\cite{CederbaumCPL1999, RemaclePNAS2006, KuleffCP2007} and experimental~\cite{CalegariScience2014} results have indicated the existence of dynamics on an attosecond timescale.

\begin{acknowledgments}
This research was supported by the Swedish Research Council, the Swedish Foundation for Strategic Research, the Knut and Alice Wallenberg Foundation, the European Research Council (PALP), the European COST Action CM1204 XLIC and the Joint Research Programme INREX of Laserlab-Europe III. 
This project has received funding from the European Union's Horizon 2020 research and innovation programme under the Marie Sklodowska-Curie grant agreement no. 641789 MEDEA.
B.F. and T.M. are financed by the project GOP-1.1.1.-12/B-2012-0001 (ELI-Hu).
\end{acknowledgments}

\hyphenation{Post-Script Sprin-ger}


\begin{thebibliography}{29}%
\makeatletter
\providecommand \@ifxundefined [1]{%
 \@ifx{#1\undefined}
}%
\providecommand \@ifnum [1]{%
 \ifnum #1\expandafter \@firstoftwo
 \else \expandafter \@secondoftwo
 \fi
}%
\providecommand \@ifx [1]{%
 \ifx #1\expandafter \@firstoftwo
 \else \expandafter \@secondoftwo
 \fi
}%
\providecommand \natexlab [1]{#1}%
\providecommand \enquote  [1]{``#1''}%
\providecommand \bibnamefont  [1]{#1}%
\providecommand \bibfnamefont [1]{#1}%
\providecommand \citenamefont [1]{#1}%
\providecommand \href@noop [0]{\@secondoftwo}%
\providecommand \href [0]{\begingroup \@sanitize@url \@href}%
\providecommand \@href[1]{\@@startlink{#1}\@@href}%
\providecommand \@@href[1]{\endgroup#1\@@endlink}%
\providecommand \@sanitize@url [0]{\catcode `\\12\catcode `\$12\catcode
  `\&12\catcode `\#12\catcode `\^12\catcode `\_12\catcode `\%12\relax}%
\providecommand \@@startlink[1]{}%
\providecommand \@@endlink[0]{}%
\providecommand \url  [0]{\begingroup\@sanitize@url \@url }%
\providecommand \@url [1]{\endgroup\@href {#1}{\urlprefix }}%
\providecommand \urlprefix  [0]{URL }%
\providecommand \Eprint [0]{\href }%
\providecommand \doibase [0]{http://dx.doi.org/}%
\providecommand \selectlanguage [0]{\@gobble}%
\providecommand \bibinfo  [0]{\@secondoftwo}%
\providecommand \bibfield  [0]{\@secondoftwo}%
\providecommand \translation [1]{[#1]}%
\providecommand \BibitemOpen [0]{}%
\providecommand \bibitemStop [0]{}%
\providecommand \bibitemNoStop [0]{.\EOS\space}%
\providecommand \EOS [0]{\spacefactor3000\relax}%
\providecommand \BibitemShut  [1]{\csname bibitem#1\endcsname}%
\let\auto@bib@innerbib\@empty
\bibitem [{\citenamefont {Knapp}\ \emph {et~al.}(2002)\citenamefont {Knapp},
  \citenamefont {Kheifets}, \citenamefont {Bray}, \citenamefont {Weber},
  \citenamefont {Landers}, \citenamefont {Sch\"ossler}, \citenamefont {Jahnke},
  \citenamefont {Nickles}, \citenamefont {Kammer}, \citenamefont {Jagutzki},
  \citenamefont {Schmidt}, \citenamefont {Osipov}, \citenamefont {R\"osch},
  \citenamefont {Prior}, \citenamefont {Schmidt-B\"ocking}, \citenamefont
  {Cocke},\ and\ \citenamefont {D\"orner}}]{KnappPRL2002}%
  \BibitemOpen
  \bibfield  {author} {\bibinfo {author} {\bibfnamefont {A.}~\bibnamefont
  {Knapp}}, \bibinfo {author} {\bibfnamefont {A.}~\bibnamefont {Kheifets}},
  \bibinfo {author} {\bibfnamefont {I.}~\bibnamefont {Bray}}, \bibinfo {author}
  {\bibfnamefont {T.}~\bibnamefont {Weber}}, \bibinfo {author} {\bibfnamefont
  {A.~L.}\ \bibnamefont {Landers}}, \bibinfo {author} {\bibfnamefont
  {S.}~\bibnamefont {Sch\"ossler}}, \bibinfo {author} {\bibfnamefont
  {T.}~\bibnamefont {Jahnke}}, \bibinfo {author} {\bibfnamefont
  {J.}~\bibnamefont {Nickles}}, \bibinfo {author} {\bibfnamefont
  {S.}~\bibnamefont {Kammer}}, \bibinfo {author} {\bibfnamefont
  {O.}~\bibnamefont {Jagutzki}}, \bibinfo {author} {\bibfnamefont {L.~P.~H.}\
  \bibnamefont {Schmidt}}, \bibinfo {author} {\bibfnamefont {T.}~\bibnamefont
  {Osipov}}, \bibinfo {author} {\bibfnamefont {J.}~\bibnamefont {R\"osch}},
  \bibinfo {author} {\bibfnamefont {M.~H.}\ \bibnamefont {Prior}}, \bibinfo
  {author} {\bibfnamefont {H.}~\bibnamefont {Schmidt-B\"ocking}}, \bibinfo
  {author} {\bibfnamefont {C.~L.}\ \bibnamefont {Cocke}}, \ and\ \bibinfo
  {author} {\bibfnamefont {R.}~\bibnamefont {D\"orner}},\ }\href {\doibase
  10.1103/PhysRevLett.89.033004} {\bibfield  {journal} {\bibinfo  {journal}
  {Phys. Rev. Lett.}\ }\textbf {\bibinfo {volume} {89}},\ \bibinfo {pages}
  {033004} (\bibinfo {year} {2002})}\BibitemShut {NoStop}%
\bibitem [{\citenamefont {Akoury}\ \emph {et~al.}(2007)\citenamefont {Akoury},
  \citenamefont {Kreidi}, \citenamefont {Jahnke}, \citenamefont {Weber},
  \citenamefont {Staudte}, \citenamefont {Schöffler}, \citenamefont {Neumann},
  \citenamefont {Titze}, \citenamefont {Schmidt}, \citenamefont {Czasch},
  \citenamefont {Jagutzki}, \citenamefont {Fraga}, \citenamefont {Grisenti},
  \citenamefont {Muiño}, \citenamefont {Cherepkov}, \citenamefont {Semenov},
  \citenamefont {Ranitovic}, \citenamefont {Cocke}, \citenamefont {Osipov},
  \citenamefont {Adaniya}, \citenamefont {Thompson}, \citenamefont {Prior},
  \citenamefont {Belkacem}, \citenamefont {Landers}, \citenamefont
  {Schmidt-B\"ocking},\ and\ \citenamefont {D\"orner}}]{AkouryScience2007}%
  \BibitemOpen
  \bibfield  {author} {\bibinfo {author} {\bibfnamefont {D.}~\bibnamefont
  {Akoury}}, \bibinfo {author} {\bibfnamefont {K.}~\bibnamefont {Kreidi}},
  \bibinfo {author} {\bibfnamefont {T.}~\bibnamefont {Jahnke}}, \bibinfo
  {author} {\bibfnamefont {T.}~\bibnamefont {Weber}}, \bibinfo {author}
  {\bibfnamefont {A.}~\bibnamefont {Staudte}}, \bibinfo {author} {\bibfnamefont
  {M.}~\bibnamefont {Schöffler}}, \bibinfo {author} {\bibfnamefont
  {N.}~\bibnamefont {Neumann}}, \bibinfo {author} {\bibfnamefont
  {J.}~\bibnamefont {Titze}}, \bibinfo {author} {\bibfnamefont {L.~P.~H.}\
  \bibnamefont {Schmidt}}, \bibinfo {author} {\bibfnamefont {A.}~\bibnamefont
  {Czasch}}, \bibinfo {author} {\bibfnamefont {O.}~\bibnamefont {Jagutzki}},
  \bibinfo {author} {\bibfnamefont {R.~A.~C.}\ \bibnamefont {Fraga}}, \bibinfo
  {author} {\bibfnamefont {R.~E.}\ \bibnamefont {Grisenti}}, \bibinfo {author}
  {\bibfnamefont {R.~D.}\ \bibnamefont {Muiño}}, \bibinfo {author}
  {\bibfnamefont {N.~A.}\ \bibnamefont {Cherepkov}}, \bibinfo {author}
  {\bibfnamefont {S.~K.}\ \bibnamefont {Semenov}}, \bibinfo {author}
  {\bibfnamefont {P.}~\bibnamefont {Ranitovic}}, \bibinfo {author}
  {\bibfnamefont {C.~L.}\ \bibnamefont {Cocke}}, \bibinfo {author}
  {\bibfnamefont {T.}~\bibnamefont {Osipov}}, \bibinfo {author} {\bibfnamefont
  {H.}~\bibnamefont {Adaniya}}, \bibinfo {author} {\bibfnamefont {J.~C.}\
  \bibnamefont {Thompson}}, \bibinfo {author} {\bibfnamefont {M.~H.}\
  \bibnamefont {Prior}}, \bibinfo {author} {\bibfnamefont {A.}~\bibnamefont
  {Belkacem}}, \bibinfo {author} {\bibfnamefont {A.~L.}\ \bibnamefont
  {Landers}}, \bibinfo {author} {\bibfnamefont {H.}~\bibnamefont
  {Schmidt-B\"ocking}}, \ and\ \bibinfo {author} {\bibfnamefont
  {R.}~\bibnamefont {D\"orner}},\ }\href {\doibase 10.1126/science.1144959}
  {\bibfield  {journal} {\bibinfo  {journal} {Science}\ }\textbf {\bibinfo
  {volume} {318}},\ \bibinfo {pages} {949} (\bibinfo {year}
  {2007})}\BibitemShut {NoStop}%
\bibitem [{\citenamefont {Agostini}\ \emph {et~al.}(1970)\citenamefont
  {Agostini}, \citenamefont {Barjot}, \citenamefont {Mainfray}, \citenamefont
  {Manus},\ and\ \citenamefont {Thebault}}]{AgostiniIEEE1970}%
  \BibitemOpen
  \bibfield  {author} {\bibinfo {author} {\bibfnamefont {P.}~\bibnamefont
  {Agostini}}, \bibinfo {author} {\bibfnamefont {G.}~\bibnamefont {Barjot}},
  \bibinfo {author} {\bibfnamefont {G.}~\bibnamefont {Mainfray}}, \bibinfo
  {author} {\bibfnamefont {C.}~\bibnamefont {Manus}}, \ and\ \bibinfo {author}
  {\bibfnamefont {J.}~\bibnamefont {Thebault}},\ }\href@noop {} {\bibfield
  {journal} {\bibinfo  {journal} {I{EEE} {J}. {Q}uant. {E}lectron.}\ }\textbf
  {\bibinfo {volume} {QE-6}},\ \bibinfo {pages} {782} (\bibinfo {year}
  {1970})}\BibitemShut {NoStop}%
\bibitem [{\citenamefont {Sorokin}\ \emph {et~al.}(2007)\citenamefont
  {Sorokin}, \citenamefont {Wellh{\"o}fer}, \citenamefont {Bobashev},
  \citenamefont {Tiedtke},\ and\ \citenamefont {Richter}}]{SorokinPRA2007}%
  \BibitemOpen
  \bibfield  {author} {\bibinfo {author} {\bibfnamefont {A. A.}~\bibnamefont
  {Sorokin}}, \bibinfo {author} {\bibfnamefont {M.}~\bibnamefont
  {Wellh{\"o}fer}}, \bibinfo {author} {\bibfnamefont {S. V.}~\bibnamefont
  {Bobashev}}, \bibinfo {author} {\bibfnamefont {K.}~\bibnamefont {Tiedtke}}, \
  and\ \bibinfo {author} {\bibfnamefont {M.}~\bibnamefont {Richter}},\ }\href
  {http://journals.aps.org/pra/abstract/10.1103/PhysRevA.75.051402} {\bibfield
  {journal} {\bibinfo  {journal} {Physical Review A}\ }\textbf {\bibinfo
  {volume} {75}},\ \bibinfo {pages} {051402} (\bibinfo {year}
  {2007})}\BibitemShut {NoStop}%
\bibitem [{\citenamefont {Papadogiannis}\ \emph {et~al.}(2003)\citenamefont
  {Papadogiannis}, \citenamefont {Nikolopoulos}, \citenamefont {Charalambidis},
  \citenamefont {Tsakiris}, \citenamefont {Tzallas},\ and\ \citenamefont
  {Witte}}]{PapadogiannisPRL2003}%
  \BibitemOpen
  \bibfield  {author} {\bibinfo {author} {\bibfnamefont {N.~A.}\ \bibnamefont
  {Papadogiannis}}, \bibinfo {author} {\bibfnamefont {L.~A.~A.}\ \bibnamefont
  {Nikolopoulos}}, \bibinfo {author} {\bibfnamefont {D.}~\bibnamefont
  {Charalambidis}}, \bibinfo {author} {\bibfnamefont {G.~D.}\ \bibnamefont
  {Tsakiris}}, \bibinfo {author} {\bibfnamefont {P.}~\bibnamefont {Tzallas}}, \
  and\ \bibinfo {author} {\bibfnamefont {K.}~\bibnamefont {Witte}},\
  }\href@noop {} {\bibfield  {journal} {\bibinfo  {journal} {Phys. {R}ev.
  {L}ett.}\ }\textbf {\bibinfo {volume} {\textbf{90}}},\ \bibinfo {pages}
  {133902} (\bibinfo {year} {2003})}\BibitemShut {NoStop}%
\bibitem [{\citenamefont {Midorikawa}\ \emph {et~al.}(2008)\citenamefont
  {Midorikawa}, \citenamefont {Nabekawa},\ and\ \citenamefont
  {Suda}}]{MidorikawaPiQE2008}%
  \BibitemOpen
  \bibfield  {author} {\bibinfo {author} {\bibfnamefont {K.}~\bibnamefont
  {Midorikawa}}, \bibinfo {author} {\bibfnamefont {Y.}~\bibnamefont
  {Nabekawa}}, \ and\ \bibinfo {author} {\bibfnamefont {A.}~\bibnamefont
  {Suda}},\ }\href@noop {} {\bibfield  {journal} {\bibinfo  {journal} {Progress
  in Quantum Electronics}\ }\textbf {\bibinfo {volume} {32}},\ \bibinfo {pages}
  {43} (\bibinfo {year} {2008})}\BibitemShut {NoStop}%
\bibitem [{\citenamefont {Moshammer}\ \emph {et~al.}(2007)\citenamefont
  {Moshammer}, \citenamefont {Jiang}, \citenamefont {Foucar}, \citenamefont
  {Rudenko}, \citenamefont {Ergler}, \citenamefont {Schr\"oter}, \citenamefont
  {L\"udemann}, \citenamefont {Zrost}, \citenamefont {Fischer}, \citenamefont
  {Titze}, \citenamefont {Jahnke}, \citenamefont {Sch\"offler}, \citenamefont
  {Weber}, \citenamefont {D\"orner}, \citenamefont {Zouros}, \citenamefont
  {Dorn}, \citenamefont {Ferger}, \citenamefont {K\"uhnel}, \citenamefont
  {D\"usterer}, \citenamefont {Treusch}, \citenamefont {Radcliffe},
  \citenamefont {Pl\"onjes},\ and\ \citenamefont {Ullrich}}]{MoshammerPRL2007}%
  \BibitemOpen
  \bibfield  {author} {\bibinfo {author} {\bibfnamefont {R.}~\bibnamefont
  {Moshammer}}, \bibinfo {author} {\bibfnamefont {Y.~H.}\ \bibnamefont
  {Jiang}}, \bibinfo {author} {\bibfnamefont {L.}~\bibnamefont {Foucar}},
  \bibinfo {author} {\bibfnamefont {A.}~\bibnamefont {Rudenko}}, \bibinfo
  {author} {\bibfnamefont {T.}~\bibnamefont {Ergler}}, \bibinfo {author}
  {\bibfnamefont {C.~D.}\ \bibnamefont {Schr\"oter}}, \bibinfo {author}
  {\bibfnamefont {S.}~\bibnamefont {L\"udemann}}, \bibinfo {author}
  {\bibfnamefont {K.}~\bibnamefont {Zrost}}, \bibinfo {author} {\bibfnamefont
  {D.}~\bibnamefont {Fischer}}, \bibinfo {author} {\bibfnamefont
  {J.}~\bibnamefont {Titze}}, \bibinfo {author} {\bibfnamefont
  {T.}~\bibnamefont {Jahnke}}, \bibinfo {author} {\bibfnamefont
  {M.}~\bibnamefont {Sch\"offler}}, \bibinfo {author} {\bibfnamefont
  {T.}~\bibnamefont {Weber}}, \bibinfo {author} {\bibfnamefont
  {R.}~\bibnamefont {D\"orner}}, \bibinfo {author} {\bibfnamefont {T.~J.~M.}\
  \bibnamefont {Zouros}}, \bibinfo {author} {\bibfnamefont {A.}~\bibnamefont
  {Dorn}}, \bibinfo {author} {\bibfnamefont {T.}~\bibnamefont {Ferger}},
  \bibinfo {author} {\bibfnamefont {K.~U.}\ \bibnamefont {K\"uhnel}}, \bibinfo
  {author} {\bibfnamefont {S.}~\bibnamefont {D\"usterer}}, \bibinfo {author}
  {\bibfnamefont {R.}~\bibnamefont {Treusch}}, \bibinfo {author} {\bibfnamefont
  {P.}~\bibnamefont {Radcliffe}}, \bibinfo {author} {\bibfnamefont
  {E.}~\bibnamefont {Pl\"onjes}}, \ and\ \bibinfo {author} {\bibfnamefont
  {J.}~\bibnamefont {Ullrich}},\ }\href {\doibase
  10.1103/PhysRevLett.98.203001} {\bibfield  {journal} {\bibinfo  {journal}
  {Phys. Rev. Lett.}\ }\textbf {\bibinfo {volume} {98}},\ \bibinfo {pages}
  {203001} (\bibinfo {year} {2007})}\BibitemShut {NoStop}%
\bibitem [{\citenamefont {Kurka}\ \emph {et~al.}(2009)\citenamefont {Kurka},
  \citenamefont {Rudenko}, \citenamefont {Foucar}, \citenamefont {K{\"u}hnel},
  \citenamefont {Jiang}, \citenamefont {Ergler}, \citenamefont {Havermeier},
  \citenamefont {Smolarski}, \citenamefont {Sch{\"o}ssler}, \citenamefont
  {Cole} \emph {et~al.}}]{KurkaJPB2009}%
  \BibitemOpen
  \bibfield  {author} {\bibinfo {author} {\bibfnamefont {M.}~\bibnamefont
  {Kurka}}, \bibinfo {author} {\bibfnamefont {A.}~\bibnamefont {Rudenko}},
  \bibinfo {author} {\bibfnamefont {L.}~\bibnamefont {Foucar}}, \bibinfo
  {author} {\bibfnamefont {K.}~\bibnamefont {K{\"u}hnel}}, \bibinfo {author}
  {\bibfnamefont {Y.}~\bibnamefont {Jiang}}, \bibinfo {author} {\bibfnamefont
  {T.}~\bibnamefont {Ergler}}, \bibinfo {author} {\bibfnamefont
  {T.}~\bibnamefont {Havermeier}}, \bibinfo {author} {\bibfnamefont
  {M.}~\bibnamefont {Smolarski}}, \bibinfo {author} {\bibfnamefont
  {S.}~\bibnamefont {Sch{\"o}ssler}}, \bibinfo {author} {\bibfnamefont
  {K.}~\bibnamefont {Cole}},  \emph {et~al.},\ }\href@noop {} {\bibfield
  {journal} {\bibinfo  {journal} {Journal of Physics B: Atomic, Molecular and
  Optical Physics}\ }\textbf {\bibinfo {volume} {42}},\ \bibinfo {pages}
  {141002} (\bibinfo {year} {2009})}\BibitemShut {NoStop}%
\bibitem [{\citenamefont {Tzallas}\ \emph {et~al.}(2003)\citenamefont
  {Tzallas}, \citenamefont {Charalambidis}, \citenamefont {Papadogiannis},
  \citenamefont {Witte},\ and\ \citenamefont {Tsakiris}}]{TzallasNature2003}%
  \BibitemOpen
  \bibfield  {author} {\bibinfo {author} {\bibfnamefont {P.}~\bibnamefont
  {Tzallas}}, \bibinfo {author} {\bibfnamefont {D.}~\bibnamefont
  {Charalambidis}}, \bibinfo {author} {\bibfnamefont {N.~A.}\ \bibnamefont
  {Papadogiannis}}, \bibinfo {author} {\bibfnamefont {K.}~\bibnamefont
  {Witte}}, \ and\ \bibinfo {author} {\bibfnamefont {G.~D.}\ \bibnamefont
  {Tsakiris}},\ }\href@noop {} {\bibfield  {journal} {\bibinfo  {journal}
  {Nature}\ }\textbf {\bibinfo {volume} {426}},\ \bibinfo {pages} {267}
  (\bibinfo {year} {2003})}\BibitemShut {NoStop}%
\bibitem [{\citenamefont {Nabekawa}\ \emph {et~al.}(2005)\citenamefont
  {Nabekawa}, \citenamefont {Hasegawa}, \citenamefont {Takahashi},\ and\
  \citenamefont {Midorikawa}}]{NabekawaPRL2005}%
  \BibitemOpen
  \bibfield  {author} {\bibinfo {author} {\bibfnamefont {Y.}~\bibnamefont
  {Nabekawa}}, \bibinfo {author} {\bibfnamefont {H.}~\bibnamefont {Hasegawa}},
  \bibinfo {author} {\bibfnamefont {E.~J.}\ \bibnamefont {Takahashi}}, \ and\
  \bibinfo {author} {\bibfnamefont {K.}~\bibnamefont {Midorikawa}},\
  }\href@noop {} {\bibfield  {journal} {\bibinfo  {journal} {Phys. {R}ev.
  {L}ett.}\ }\textbf {\bibinfo {volume} {94}},\ \bibinfo {pages} {043001}
  (\bibinfo {year} {2005})}\BibitemShut {NoStop}%
\bibitem [{\citenamefont {Takahashi}\ \emph {et~al.}(2013)\citenamefont
  {Takahashi}, \citenamefont {Lan}, \citenamefont {M\"ucke}, \citenamefont
  {Nabekawa},\ and\ \citenamefont {Midorikawa}}]{TakahashiNC2013}%
  \BibitemOpen
  \bibfield  {author} {\bibinfo {author} {\bibfnamefont {E.~J.}\ \bibnamefont
  {Takahashi}}, \bibinfo {author} {\bibfnamefont {P.}~\bibnamefont {Lan}},
  \bibinfo {author} {\bibfnamefont {O.~D.}\ \bibnamefont {M\"ucke}}, \bibinfo
  {author} {\bibfnamefont {Y.}~\bibnamefont {Nabekawa}}, \ and\ \bibinfo
  {author} {\bibfnamefont {K.}~\bibnamefont {Midorikawa}},\ }\href@noop {}
  {\bibfield  {journal} {\bibinfo  {journal} {Nature Communications}\ }\textbf
  {\bibinfo {volume} {4}} (\bibinfo {year} {2013})}\BibitemShut {NoStop}%
\bibitem [{\citenamefont {Tzallas}\ \emph {et~al.}(2011)\citenamefont
  {Tzallas}, \citenamefont {Skantzakis}, \citenamefont {Nikolopoulos},
  \citenamefont {Tsakiris},\ and\ \citenamefont
  {Charalambidis}}]{TzallasNP2011}%
  \BibitemOpen
  \bibfield  {author} {\bibinfo {author} {\bibfnamefont {P.}~\bibnamefont
  {Tzallas}}, \bibinfo {author} {\bibfnamefont {E.}~\bibnamefont {Skantzakis}},
  \bibinfo {author} {\bibfnamefont {L.}~\bibnamefont {Nikolopoulos}}, \bibinfo
  {author} {\bibfnamefont {G.}~\bibnamefont {Tsakiris}}, \ and\ \bibinfo
  {author} {\bibfnamefont {D.}~\bibnamefont {Charalambidis}},\ }\href@noop {}
  {\bibfield  {journal} {\bibinfo  {journal} {Nature Physics}\ }\textbf
  {\bibinfo {volume} {7}},\ \bibinfo {pages} {781} (\bibinfo {year}
  {2011})}\BibitemShut {NoStop}%
\bibitem [{\citenamefont {Covington}\ \emph {et~al.}(2002)\citenamefont
  {Covington}, \citenamefont {Aguilar}, \citenamefont {Covington},
  \citenamefont {Gharaibeh}, \citenamefont {Hinojosa}, \citenamefont {Shirley},
  \citenamefont {Phaneuf}, \citenamefont {Alvarez}, \citenamefont {Cisneros},
  \citenamefont {Dominguez-Lopez} \emph {et~al.}}]{CovingtonPRA2002}%
  \BibitemOpen
  \bibfield  {author} {\bibinfo {author} {\bibfnamefont {A.}~\bibnamefont
  {Covington}}, \bibinfo {author} {\bibfnamefont {A.}~\bibnamefont {Aguilar}},
  \bibinfo {author} {\bibfnamefont {I.}~\bibnamefont {Covington}}, \bibinfo
  {author} {\bibfnamefont {M.}~\bibnamefont {Gharaibeh}}, \bibinfo {author}
  {\bibfnamefont {G.}~\bibnamefont {Hinojosa}}, \bibinfo {author}
  {\bibfnamefont {C.}~\bibnamefont {Shirley}}, \bibinfo {author} {\bibfnamefont
  {R.}~\bibnamefont {Phaneuf}}, \bibinfo {author} {\bibfnamefont
  {I.}~\bibnamefont {Alvarez}}, \bibinfo {author} {\bibfnamefont
  {C.}~\bibnamefont {Cisneros}}, \bibinfo {author} {\bibfnamefont
  {I.}~\bibnamefont {Dominguez-Lopez}},  \emph {et~al.},\ }\href@noop {}
  {\bibfield  {journal} {\bibinfo  {journal} {Physical Review A}\ }\textbf
  {\bibinfo {volume} {66}},\ \bibinfo {pages} {062710} (\bibinfo {year}
  {2002})}\BibitemShut {NoStop}%
\bibitem [{\citenamefont {Feist}\ \emph {et~al.}(2009)\citenamefont {Feist},
  \citenamefont {Nagele}, \citenamefont {Pazourek}, \citenamefont {Persson},
  \citenamefont {Schneider}, \citenamefont {Collins},\ and\ \citenamefont
  {Burgd\"orfer}}]{FeistPRL2009}%
  \BibitemOpen
  \bibfield  {author} {\bibinfo {author} {\bibfnamefont {J.}~\bibnamefont
  {Feist}}, \bibinfo {author} {\bibfnamefont {S.}~\bibnamefont {Nagele}},
  \bibinfo {author} {\bibfnamefont {R.}~\bibnamefont {Pazourek}}, \bibinfo
  {author} {\bibfnamefont {E.}~\bibnamefont {Persson}}, \bibinfo {author}
  {\bibfnamefont {B.~I.}\ \bibnamefont {Schneider}}, \bibinfo {author}
  {\bibfnamefont {L.~A.}\ \bibnamefont {Collins}}, \ and\ \bibinfo {author}
  {\bibfnamefont {J.}~\bibnamefont {Burgd\"orfer}},\ }\href@noop {} {\bibfield
  {journal} {\bibinfo  {journal} {Phys. Rev. Lett.}\ }\textbf {\bibinfo
  {volume} {103}},\ \bibinfo {pages} {063002} (\bibinfo {year}
  {2009})}\BibitemShut {NoStop}%
\bibitem [{\citenamefont {Fritzsche}\ \emph {et~al.}(2008)\citenamefont
  {Fritzsche}, \citenamefont {Grum-Grzhimailo}, \citenamefont {Gryzlova},\ and\
  \citenamefont {Kabachnik}}]{FritzscheJPB2008}%
  \BibitemOpen
  \bibfield  {author} {\bibinfo {author} {\bibfnamefont {S.}~\bibnamefont
  {Fritzsche}}, \bibinfo {author} {\bibfnamefont {A.}~\bibnamefont
  {Grum-Grzhimailo}}, \bibinfo {author} {\bibfnamefont {E.}~\bibnamefont
  {Gryzlova}}, \ and\ \bibinfo {author} {\bibfnamefont {N.}~\bibnamefont
  {Kabachnik}},\ }\href@noop {} {\bibfield  {journal} {\bibinfo  {journal}
  {Journal of Physics B: Atomic, Molecular and Optical Physics}\ }\textbf
  {\bibinfo {volume} {41}},\ \bibinfo {pages} {165601} (\bibinfo {year}
  {2008})}\BibitemShut {NoStop}%
\bibitem [{\citenamefont {West}\ and\ \citenamefont
  {Marr}(1976)}]{WestPRSA1976}%
  \BibitemOpen
  \bibfield  {author} {\bibinfo {author} {\bibfnamefont {J.}~\bibnamefont
  {West}}\ and\ \bibinfo {author} {\bibfnamefont {G.}~\bibnamefont {Marr}},\
  }in\ \href@noop {} {\emph {\bibinfo {booktitle} {Proceedings of the Royal
  Society of London A: Mathematical, Physical and Engineering Sciences}}},\
  Vol.\ \bibinfo {volume} {349}\ (\bibinfo {organization} {The Royal Society},\
  \bibinfo {year} {1976})\ pp.\ \bibinfo {pages} {397--421}\BibitemShut
  {NoStop}%
\bibitem [{\citenamefont {F\o{}rre}\ \emph {et~al.}(2010)\citenamefont
  {F\o{}rre}, \citenamefont {Selst\o{}},\ and\ \citenamefont
  {Nepstad}}]{ForrePRL2010}%
  \BibitemOpen
  \bibfield  {author} {\bibinfo {author} {\bibfnamefont {M.}~\bibnamefont
  {F\o{}rre}}, \bibinfo {author} {\bibfnamefont {S.}~\bibnamefont {Selst\o{}}},
  \ and\ \bibinfo {author} {\bibfnamefont {R.}~\bibnamefont {Nepstad}},\ }\href
  {\doibase 10.1103/PhysRevLett.105.163001} {\bibfield  {journal} {\bibinfo
  {journal} {Phys. Rev. Lett.}\ }\textbf {\bibinfo {volume} {105}},\ \bibinfo
  {pages} {163001} (\bibinfo {year} {2010})}\BibitemShut {NoStop}%
\bibitem [{\citenamefont {Takahashi}\ \emph {et~al.}(2002)\citenamefont
  {Takahashi}, \citenamefont {Nabekawa}, \citenamefont {Otsuka}, \citenamefont
  {Obara},\ and\ \citenamefont {Midorikawa}}]{TakahashiPRA2002}%
  \BibitemOpen
  \bibfield  {author} {\bibinfo {author} {\bibfnamefont {E.}~\bibnamefont
  {Takahashi}}, \bibinfo {author} {\bibfnamefont {Y.}~\bibnamefont {Nabekawa}},
  \bibinfo {author} {\bibfnamefont {T.}~\bibnamefont {Otsuka}}, \bibinfo
  {author} {\bibfnamefont {M.}~\bibnamefont {Obara}}, \ and\ \bibinfo {author}
  {\bibfnamefont {K.}~\bibnamefont {Midorikawa}},\ }\href
  {http://link.aps.org/doi/10.1103/PhysRevA.66.021802} {\bibfield  {journal}
  {\bibinfo  {journal} {Phys. {R}ev. {A}}\ }\textbf {\bibinfo {volume} {66}},\
  \bibinfo {pages} {021802} (\bibinfo {year} {2002})}\BibitemShut {NoStop}%
\bibitem [{\citenamefont {Rudawski}\ \emph {et~al.}(2013)\citenamefont
  {Rudawski}, \citenamefont {Heyl}, \citenamefont {Brizuela}, \citenamefont
  {Schwenke}, \citenamefont {Persson}, \citenamefont {Mansten}, \citenamefont
  {Rakowski}, \citenamefont {Rading}, \citenamefont {Campi}, \citenamefont
  {Kim}, \citenamefont {Johnsson},\ and\ \citenamefont
  {L'Huillier}}]{RudawskiRSI2013}%
  \BibitemOpen
  \bibfield  {author} {\bibinfo {author} {\bibfnamefont {P.}~\bibnamefont
  {Rudawski}}, \bibinfo {author} {\bibfnamefont {C.~M.}\ \bibnamefont {Heyl}},
  \bibinfo {author} {\bibfnamefont {F.}~\bibnamefont {Brizuela}}, \bibinfo
  {author} {\bibfnamefont {J.}~\bibnamefont {Schwenke}}, \bibinfo {author}
  {\bibfnamefont {A.}~\bibnamefont {Persson}}, \bibinfo {author} {\bibfnamefont
  {E.}~\bibnamefont {Mansten}}, \bibinfo {author} {\bibfnamefont
  {R.}~\bibnamefont {Rakowski}}, \bibinfo {author} {\bibfnamefont
  {L.}~\bibnamefont {Rading}}, \bibinfo {author} {\bibfnamefont
  {F.}~\bibnamefont {Campi}}, \bibinfo {author} {\bibfnamefont
  {B.}~\bibnamefont {Kim}}, \bibinfo {author} {\bibfnamefont {P.}~\bibnamefont
  {Johnsson}}, \ and\ \bibinfo {author} {\bibfnamefont {A.}~\bibnamefont
  {L'Huillier}},\ }\href {\doibase http://dx.doi.org/10.1063/1.4812266}
  {\bibfield  {journal} {\bibinfo  {journal} {Review of Scientific
  Instruments}\ }\textbf {\bibinfo {volume} {84}},\ \bibinfo {eid} {073103}
  (\bibinfo {year} {2013})}\BibitemShut {NoStop}%
\bibitem [{\citenamefont {{L\'opez-Martens}}\ \emph {et~al.}(2005)\citenamefont
  {{L\'opez-Martens}}, \citenamefont {Varj\'u}, \citenamefont {Johnsson},
  \citenamefont {Mauritsson}, \citenamefont {Mairesse}, \citenamefont
  {Sali\`eres}, \citenamefont {Gaarde}, \citenamefont {Schafer}, \citenamefont
  {Persson}, \citenamefont {Svanberg}, \citenamefont {Wahlstr\"om},\ and\
  \citenamefont {L'Huillier}}]{LopezMartensPRL2005}%
  \BibitemOpen
  \bibfield  {author} {\bibinfo {author} {\bibfnamefont {R.}~\bibnamefont
  {{L\'opez-Martens}}}, \bibinfo {author} {\bibfnamefont {K.}~\bibnamefont
  {Varj\'u}}, \bibinfo {author} {\bibfnamefont {P.}~\bibnamefont {Johnsson}},
  \bibinfo {author} {\bibfnamefont {J.}~\bibnamefont {Mauritsson}}, \bibinfo
  {author} {\bibfnamefont {Y.}~\bibnamefont {Mairesse}}, \bibinfo {author}
  {\bibfnamefont {P.}~\bibnamefont {Sali\`eres}}, \bibinfo {author}
  {\bibfnamefont {M.~B.}\ \bibnamefont {Gaarde}}, \bibinfo {author}
  {\bibfnamefont {K.~J.}\ \bibnamefont {Schafer}}, \bibinfo {author}
  {\bibfnamefont {A.}~\bibnamefont {Persson}}, \bibinfo {author} {\bibfnamefont
  {S.}~\bibnamefont {Svanberg}}, \bibinfo {author} {\bibfnamefont {C.-G.}\
  \bibnamefont {Wahlstr\"om}}, \ and\ \bibinfo {author} {\bibfnamefont
  {A.}~\bibnamefont {L'Huillier}},\ }\href@noop {} {\bibfield  {journal}
  {\bibinfo  {journal} {Phys. {R}ev. {L}ett.}\ }\textbf {\bibinfo {volume}
  {94}},\ \bibinfo {pages} {033001} (\bibinfo {year} {2005})}\BibitemShut
  {NoStop}%
\bibitem [{\citenamefont {Varj\'u}\ \emph {et~al.}(2005)\citenamefont
  {Varj\'u}, \citenamefont {Mairesse}, \citenamefont {Carre}, \citenamefont
  {Gaarde}, \citenamefont {Johnsson}, \citenamefont {Kazamias}, \citenamefont
  {Lopez-Martens}, \citenamefont {Mauritsson}, \citenamefont {Schafer},
  \citenamefont {Balcou}, \citenamefont {L'Huillier},\ and\ \citenamefont
  {Sali\`eres}}]{VarjuJMO2005}%
  \BibitemOpen
  \bibfield  {author} {\bibinfo {author} {\bibfnamefont {K.}~\bibnamefont
  {Varj\'u}}, \bibinfo {author} {\bibfnamefont {Y.}~\bibnamefont {Mairesse}},
  \bibinfo {author} {\bibfnamefont {B.}~\bibnamefont {Carre}}, \bibinfo
  {author} {\bibfnamefont {M.~B.}\ \bibnamefont {Gaarde}}, \bibinfo {author}
  {\bibfnamefont {P.}~\bibnamefont {Johnsson}}, \bibinfo {author}
  {\bibfnamefont {S.}~\bibnamefont {Kazamias}}, \bibinfo {author}
  {\bibfnamefont {R.}~\bibnamefont {Lopez-Martens}}, \bibinfo {author}
  {\bibfnamefont {J.}~\bibnamefont {Mauritsson}}, \bibinfo {author}
  {\bibfnamefont {K.~J.}\ \bibnamefont {Schafer}}, \bibinfo {author}
  {\bibfnamefont {P.}~\bibnamefont {Balcou}}, \bibinfo {author} {\bibfnamefont
  {A.}~\bibnamefont {L'Huillier}}, \ and\ \bibinfo {author} {\bibfnamefont
  {P.}~\bibnamefont {Sali\`eres}},\ }\href@noop {} {\bibfield  {journal}
  {\bibinfo  {journal} {J. {M}od. {O}pt.}\ }\textbf {\bibinfo {volume} {52}},\
  \bibinfo {pages} {379} (\bibinfo {year} {2005})}\BibitemShut {NoStop}%
\bibitem [{\citenamefont {Wolter}(1952)}]{WolterADP1952}%
  \BibitemOpen
  \bibfield  {author} {\bibinfo {author} {\bibfnamefont {H.}~\bibnamefont
  {Wolter}},\ }\href@noop {} {\bibfield  {journal} {\bibinfo  {journal}
  {Annalen der Physik}\ }\textbf {\bibinfo {volume} {445}},\ \bibinfo {pages}
  {94} (\bibinfo {year} {1952})}\BibitemShut {NoStop}%
\bibitem [{\citenamefont {Wiley}\ and\ \citenamefont
  {McLaren}(1955)}]{WileyRSI1955}%
  \BibitemOpen
  \bibfield  {author} {\bibinfo {author} {\bibfnamefont {W.~C.}\ \bibnamefont
  {Wiley}}\ and\ \bibinfo {author} {\bibfnamefont {I.~H.}\ \bibnamefont
  {McLaren}},\ }\href@noop {} {\bibfield  {journal} {\bibinfo  {journal} {Rev.
  Sci. Instrum.}\ }\textbf {\bibinfo {volume} {26}},\ \bibinfo {pages} {1150}
  (\bibinfo {year} {1955})}\BibitemShut {NoStop}%
\bibitem [{\citenamefont {Lias}(2014)}]{Lias2014}%
  \BibitemOpen
  \bibfield  {author} {\bibinfo {author} {\bibfnamefont {S.}~\bibnamefont
  {Lias}},\ }\href@noop {} {\emph {\bibinfo {title} {Ionization Energy
  Evaluation}}},\ edited by\ \bibinfo {editor} {\bibfnamefont {P.}~\bibnamefont
  {Linstrom}}\ and\ \bibinfo {editor} {\bibfnamefont {W.}~\bibnamefont
  {Mallard}},\ Vol.\ \bibinfo {volume} {NIST Standard Reference Database Number
  69}\ (\bibinfo  {publisher} {NIST Chemistry WebBook},\ \bibinfo {address}
  {National Institute of Standards and Technology, Gaithersburg MD, 20899},\
  \bibinfo {year} {2014})\BibitemShut {NoStop}%
\bibitem [{\citenamefont {Palacios}\ \emph {et~al.}(2009)\citenamefont
  {Palacios}, \citenamefont {Rescigno},\ and\ \citenamefont
  {McCurdy}}]{PalaciosPRA2009}%
  \BibitemOpen
  \bibfield  {author} {\bibinfo {author} {\bibfnamefont {A.}~\bibnamefont
  {Palacios}}, \bibinfo {author} {\bibfnamefont {T.~N.}\ \bibnamefont
  {Rescigno}}, \ and\ \bibinfo {author} {\bibfnamefont {C.~W.}\ \bibnamefont
  {McCurdy}},\ }\href {\doibase 10.1103/PhysRevA.79.033402} {\bibfield
  {journal} {\bibinfo  {journal} {Phys. Rev. A}\ }\textbf {\bibinfo {volume}
  {79}},\ \bibinfo {pages} {033402} (\bibinfo {year} {2009})}\BibitemShut
  {NoStop}%
\bibitem [{\citenamefont {Cederbaum}\ and\ \citenamefont
  {Zobeley}(1999)}]{CederbaumCPL1999}%
  \BibitemOpen
  \bibfield  {author} {\bibinfo {author} {\bibfnamefont {L.~S.}\ \bibnamefont
  {Cederbaum}}\ and\ \bibinfo {author} {\bibfnamefont {J.}~\bibnamefont
  {Zobeley}},\ }\href@noop {} {\bibfield  {journal} {\bibinfo  {journal} {Chem.
  Phys. Lett.}\ }\textbf {\bibinfo {volume} {307}},\ \bibinfo {pages} {205}
  (\bibinfo {year} {1999})}\BibitemShut {NoStop}%
\bibitem [{\citenamefont {Remacle}\ and\ \citenamefont
  {Levine}(2006)}]{RemaclePNAS2006}%
  \BibitemOpen
  \bibfield  {author} {\bibinfo {author} {\bibfnamefont {F.}~\bibnamefont
  {Remacle}}\ and\ \bibinfo {author} {\bibfnamefont {R.~D.}\ \bibnamefont
  {Levine}},\ }\href@noop {} {\bibfield  {journal} {\bibinfo  {journal}
  {P{NAS}}\ }\textbf {\bibinfo {volume} {103}},\ \bibinfo {pages} {6793}
  (\bibinfo {year} {2006})}\BibitemShut {NoStop}%
\bibitem [{\citenamefont {Kuleff}\ and\ \citenamefont
  {Cederbaum}(2007)}]{KuleffCP2007}%
  \BibitemOpen
  \bibfield  {author} {\bibinfo {author} {\bibfnamefont {A.~I.}\ \bibnamefont
  {Kuleff}}\ and\ \bibinfo {author} {\bibfnamefont {L.~S.}\ \bibnamefont
  {Cederbaum}},\ }\href@noop {} {\bibfield  {journal} {\bibinfo  {journal}
  {Chem. Phys.}\ }\textbf {\bibinfo {volume} {338}},\ \bibinfo {pages} {320}
  (\bibinfo {year} {2007})}\BibitemShut {NoStop}%
\bibitem [{\citenamefont {Calegari}\ \emph {et~al.}(2014)\citenamefont
  {Calegari}, \citenamefont {Ayuso}, \citenamefont {Trabattoni}, \citenamefont
  {Belshaw}, \citenamefont {De~Camillis}, \citenamefont {Anumula},
  \citenamefont {Frassetto}, \citenamefont {Poletto}, \citenamefont {Palacios},
  \citenamefont {Decleva}, \citenamefont {Greenwood}, \citenamefont {Martin},\
  and\ \citenamefont {Nisoli}}]{CalegariScience2014}%
  \BibitemOpen
  \bibfield  {author} {\bibinfo {author} {\bibfnamefont {F.}~\bibnamefont
  {Calegari}}, \bibinfo {author} {\bibfnamefont {D.}~\bibnamefont {Ayuso}},
  \bibinfo {author} {\bibfnamefont {A.}~\bibnamefont {Trabattoni}}, \bibinfo
  {author} {\bibfnamefont {L.}~\bibnamefont {Belshaw}}, \bibinfo {author}
  {\bibfnamefont {S.}~\bibnamefont {De~Camillis}}, \bibinfo {author}
  {\bibfnamefont {S.}~\bibnamefont {Anumula}}, \bibinfo {author} {\bibfnamefont
  {F.}~\bibnamefont {Frassetto}}, \bibinfo {author} {\bibfnamefont
  {L.}~\bibnamefont {Poletto}}, \bibinfo {author} {\bibfnamefont
  {A.}~\bibnamefont {Palacios}}, \bibinfo {author} {\bibfnamefont
  {P.}~\bibnamefont {Decleva}}, \bibinfo {author} {\bibfnamefont {J.~B.}\
  \bibnamefont {Greenwood}}, \bibinfo {author} {\bibfnamefont {F.}~\bibnamefont
  {Martin}}, \ and\ \bibinfo {author} {\bibfnamefont {M.}~\bibnamefont
  {Nisoli}},\ }\href@noop {} {\bibfield  {journal} {\bibinfo  {journal}
  {Science}\ }\textbf {\bibinfo {volume} {346}},\ \bibinfo {pages} {336}
  (\bibinfo {year} {2014})}\BibitemShut {NoStop}%
\end{thebibliography}
\end{document}